\documentclass[aps,twocolumn,nofootinbib,prl]{revtex4-1}

\usepackage{bm, amsmath,amssymb,amsfonts,slashed,array,graphicx,soul}
\usepackage[vcentermath,noautoscale]{youngtab}
\DeclareMathAlphabet{\mathpzc}{OT1}{pzc}{m}{it}

\Yboxdim{8pt}
\newcommand{\fund}{\yng(1)}
\newcommand{\afund}{\overline{\yng(1)}}

\newcommand{\UV}{U(1)_{\rm{V}}}
\newcommand{\UY}{U(1)_{\rm{Y}}}
\newcommand{\SUV}{SU(N)_{\rm{V}}}

\newcommand{\id}{\bm{1}}
\newcommand{\etav}{\eta_v'}
\newcommand{\uv}{\mathpzc{u}}
\newcommand{\bv}{\mathpzc{b}}
\newcommand{\qv}{\mathpzc{q}}
\newcommand{\uL}{\uv_L^{\phantom{c}}}
\newcommand{\bL}{\bv_L^{\phantom{c}}}
\newcommand{\uRc}{\uv_R^{c}}
\newcommand{\bRc}{\bv_R^{c}}
\newcommand{\xL}{{\lambda_1}_L^{\phantom{c}}}
\newcommand{\zL}{{\lambda_2}_L^{\phantom{c}}}
\newcommand{\xRc}{{\lambda_1}_R^{c}}
\newcommand{\zRc}{{\lambda_2}_R^{c}}
\newcommand{\Lambdav}{\Lambda_v}
\newcommand{\sigmav}{\sigma_v}

\usepackage{xcolor}
\definecolor{red}{rgb}{0.9, 0,0}

\usepackage{hyperref}	
\begin{document}

\title{Galactic Center Gamma-Ray Excess through a Dark Shower}
\author{Marat Freytsis}
\affiliation{Department of Physics, Harvard University, Cambridge, MA 02138}
\author{Dean J. Robinson}
\affiliation{Department of Physics, University of California, Berkeley, CA 94720, USA}
\affiliation{Ernest Orlando Lawrence Berkeley National Laboratory,
University of California, Berkeley, CA 94720, USA}
\author{Yuhsin Tsai}
\affiliation{Department of Physics, University of California, Davis, CA 95616, USA}
\begin{abstract}
The reported excess of $\gamma$-rays, emitted from an extended region around the galactic center, has a distribution and rate suggestive of an origin in dark matter (DM) annihilations. The conventional annihilation channels into standard model (SM) $b$ quarks or $\tau$ leptons may, however, be in tension with various experimental constraints on antiproton and positron fluxes. We present a framework that is free from such constraints. The key idea is that the mediators between the dark matter and the SM are themselves part of a strongly coupled sector: a hidden valley. In this scenario, the dark matter particles annihilate only into hidden quarks that subsequently shower and hadronize. Hidden quark effective couplings to SM hypercharge allow the lightest hidden bound states to subsequently decay into SM photons, producing the observed photon energy spectrum. Associated production of SM fermions is, in contrast, suppressed by electroweak, loop or helicity effects. We find that, generically, $\sim10$ GeV DM and a confinement scale $\sim 1$~GeV provide a good fit to the observed spectrum. An $SU(2)$ hidden confining group is preferred over $SU(3)$ or higher rank gauge groups, up to uncertainties in the extraction of the astrophysical background. An explicit realization of this framework is also presented, and its phenomenology is discussed in detail, along with pertinent cosmological, astrophysical and collider bounds. This framework may be probed by model-independent searches, including future beam-dump experiments.

\end{abstract}
\maketitle

\textbf{Introduction.}  Recent results from \emph{Fermi}-LAT have reconfirmed reports of a GeV gamma-ray signal originating from central regions of the Milky Way galaxy, with very high statistical significance \cite{Goodenough:2009gk,Hooper:2010mq,Hooper:2011og,Abazajian:2012dgr,Abazajian:2012edg,Gordon:2013vta,Hooper:2013rwa,Daylan:2014rsa,Abazajian:2014fta,Calore:2014xka}. The energy spectrum of this signal closely resembles that of $\gamma$-rays originating in a parton shower, while the radial spectrum is spherically symmetric with a halo profile $\sim r^{-\gamma}$, $\gamma \simeq1.1$--$1.3$ \cite{Daylan:2014rsa}. Barring possible astrophysical sources for this signal \cite{Baltz:2006sv,Abazajian:2012dgr,Abazajian:2012edg,Linden:2012tm,Gordon:2013vta,Yusef-Zadeh:2013:ic,Cholis:2014lta,Cholis:2014noa}, such as millisecond pulsars, a hypothetical source of this GeV signal is the annihilation of GeV-scale dark matter (DM) into Standard Model (SM) particles. With the recent \emph{Fermi}-LAT data, it has been shown that the signal is well-matched by the annihilation of weakly coupled 30-40 GeV DM into $b$ quarks or 7-10 GeV DM into $\tau$ leptons. The required cross-section is of the appropriate scale for the DM to form a cold dark matter thermal relic (see e.g. Refs \cite{Hooper:2010mq,Abazajian:2012dgr,Daylan:2014rsa,Berlin:2014pya}). 

This typical scenario -- GeV DM annihilation into SM quarks and/or leptons, respectively the `hadronic' and `leptonic' scenarios -- can be produced in many different extensions of the SM (see e.g. Refs \cite{Abdullah:2014lla,Agrawal:2014una,Alves:2014yha,Berlin:2014pya,Berlin:2014tja,Boehm:2014bia,Cerdeno:2014cda,Cheung:2014lqa,Cline:2014dwa,Ghosh:2014pwa,Huang:2014cla,Ipek:2014gua,Izaguirre:2014vva,Ko:2014gha,Martin:2014sxa} for recent studies). However, such DM annihilations also involve the associated emission of cosmic rays, in particular antiprotons and positrons, and inverse Compton scattering or bremsstrahlung radiation. Current antiproton and positron data, together with radio observations, may either strongly constrain or is in tension with the hadronic and leptonic scenarios \cite{Bertone:2001jv, Regis:2008ij, Bringmann:2009ca,Modak:2013jya,Ng:2013xha,Cirelli:2014lwa,Pettorino:2014sua,Bringmann:2014lpa,Detmold:2014qqa,Cholis:2014fja,Hooper:2014ysa, Bell:2014xta, Arina:2014yna, Cao:2014efa}, up to uncertainties regarding galactic propagation.

\begin{figure}[h]
	\includegraphics[width = 6cm]{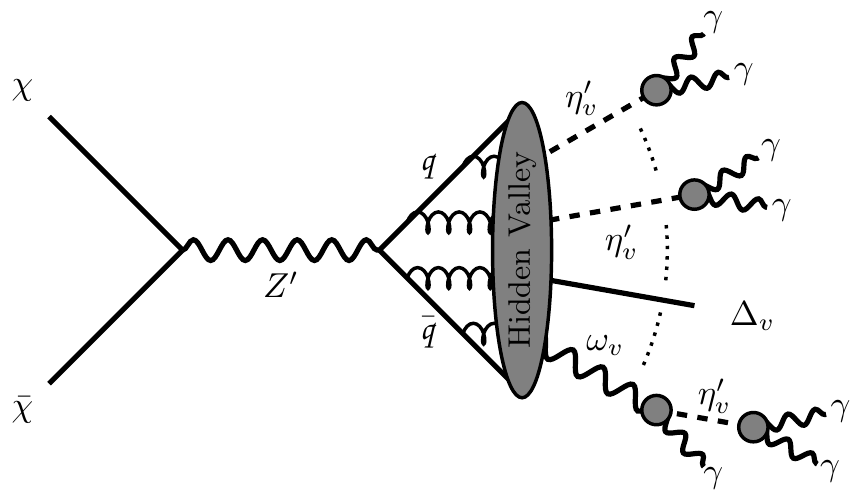}
	\caption{Schematic amplitude for GeV DM, $\chi$, annihilation through a hidden $Z'$ to hidden quarks, $\qv$, followed by showering to hidden pseudoscalars, $\etav$. The shower may include other light (meta)stable bound states, $\omega_v$ and $\Delta_v$ (defined below). The $\etav$'s subsequently dominantly decay to photons, as may other metastable states.}
	\label{fig:DA}
\end{figure}

In this work, we therefore propose an alternative picture: We imagine that SM sterile dark matter instead couples weakly to a hidden, strongly coupled sector -- a hidden valley (HV) \cite{Strassler:2006im,Han:2007ae} (see also e.g. Refs \cite{Cassel:2009pu,Dubovsky:2010je,Krolikowski:2008qa,Strassler:2008bv}) -- that itself has effective millicharged couplings to the SM. That is, the DM does not annihilate directly into SM quarks or leptons, which subsequently shower into SM hadrons, leptons and photons. Rather, in this picture the DM instead annihilates into HV quarks, that shower predominantly into light hidden valley pseudoscalars. These pseudoscalars subsequently dominantly decay through a millicharged chiral anomaly into photons, producing the observed parton-shower-like prompt photon spectrum. Associated SM matter production from hidden pseudoscalar decay, such as electron-positron or proton-antiproton production, may only occur through Dalitz or higher loop electroweak processes, and is correspondingly further suppressed or may even be kinematically forbidden, depending on the pseudoscalar mass. (See Fig.~\ref{fig:DA} for a schematic description of this showering framework.) The prompt photon spectrum is therefore produced without significant associated antiproton or positron production. In this fashion, the aforementioned constraints or tensions are duly lifted from the DM annihilation scenario.

\textbf{Galactic center (GC) $\gamma$-ray excess simulations.}
We first illustrate the ability of this general HV scenario to produce the observed excess of $\gamma$-rays from the GC. That
this is possible should not come as a great surprise. Just as in the annihilation of
DM to quarks -- already shown to provide a good fit to the signal --  the energy
distribution of photons is primarily controlled by the structure of the parton
shower. Although the details of the spectrum of bound states may be different for the HV scenario,
its gauge group and coupling strength are also unfixed, allowing these
to compensate for differences due to hadronization.

We simulate $\gamma$-ray spectra for different choices of $m_\chi$, the DM mass,
and $\Lambdav$, the Landau pole of the HV confining group, an analogue of
$\Lambda_\text{QCD}$ for the gauge group $\SUV$. The spectra are generated by
\textsc{Pythia8}~\cite{Sjostrand:2007gs}, modified to include the one-loop
running of the hidden valley gauge group. This is necessary to correctly capture
the resulting photon energy spectrum. Computation of astrophysical rates 
is done using \textsc{PPPC 4 DM ID}~\cite{Cirelli:2010xx}. For the HV models
we consider (defined in detail below), after setting the value of $N$,
the choice of $m_\chi$ and $\Lambdav$ fixes all the remaining
HV parameters in the confining sector required to model the parton shower.

After hadronization, we only model the lightest (meta)stable bound states. Working
with a one-flavor sector, these are taken to be a pseudoscalar decaying as $\etav
\to \gamma\gamma$, a vector decaying as $\omega_v \to \etav \gamma$ and a stable
baryon, $\Delta_v$, which can be a fermion or boson depending on the choice
of confining gauge group. The photon production rate is therefore sensitive to the ratio
of baryons to mesons after hadronization. For $SU(3)_\text{V}$, we take this fraction
to be $\mathcal{O}(10\%)$, following QCD, and assume a negligible fraction for larger gauge groups. For $SU(2)_\text{V}$, however, we expect no suppression from the number of colors. The relative fraction is then determined purely by masses and spin.
Using production rates of light hadrons at LEP~\cite{Chliapnikov:1999qi}
to set approximate suppression rates arising from mass effects, and assuming no mass
splitting for $\omega_v$ and $\Delta_v$, we estimate the baryon fraction to be
$\mathcal{O}(35\%)$.

The spectra are also sensitive to the underlying hadronization model, which we estimate to introduce
an $\mathcal{O}(10\%)$ systematic uncertainty in the goodness-of-fit
at the best-fit point, with the uncertainty reducing for increased $m_\chi/\Lambdav$.
No attempt has been made here to tune the default hadronization behavior
of the \textsc{Pythia8} hidden valley hadronization model.

A full simulation of the GC excess signal plus background for different DM annihilation and showering templates is beyond the scope of this work. While the location and shape of the spectral peak is expected to be robust under variation of the DM template \cite{Abazajian:2014fta,Daylan:2014rsa}, the length of the low and high energy photon tails in the signal spectrum  may vary as the background best-fits fluctuate. The various signal templates considered in Ref.~\cite{Daylan:2014rsa}, however, imply little variation in the extracted signal once showering kinematics are fixed. We therefore assume hereafter that the background is the same as for the $b\bar{b}$ DM template presented in Ref.~\cite{Daylan:2014rsa}: The simulated spectra are then fit against the signal data reported therein. We treat possible variations in the tails as inherent systematic uncertainties.

The fits themselves are performed by allowing the annihilation cross-section to float at each point in $m_\chi$-$\Lambdav$ space for $N=2,3,4$, and minimizing the resulting value of Pearson's $\chi^2$ statistic.  Because of unquantified and unquantifiable systematic errors associated with these spectra, we do not attempt to assign statistical significance to these fits, but only measure goodness-of-fit.

\begin{figure}[t]
	\includegraphics[width = 7cm]{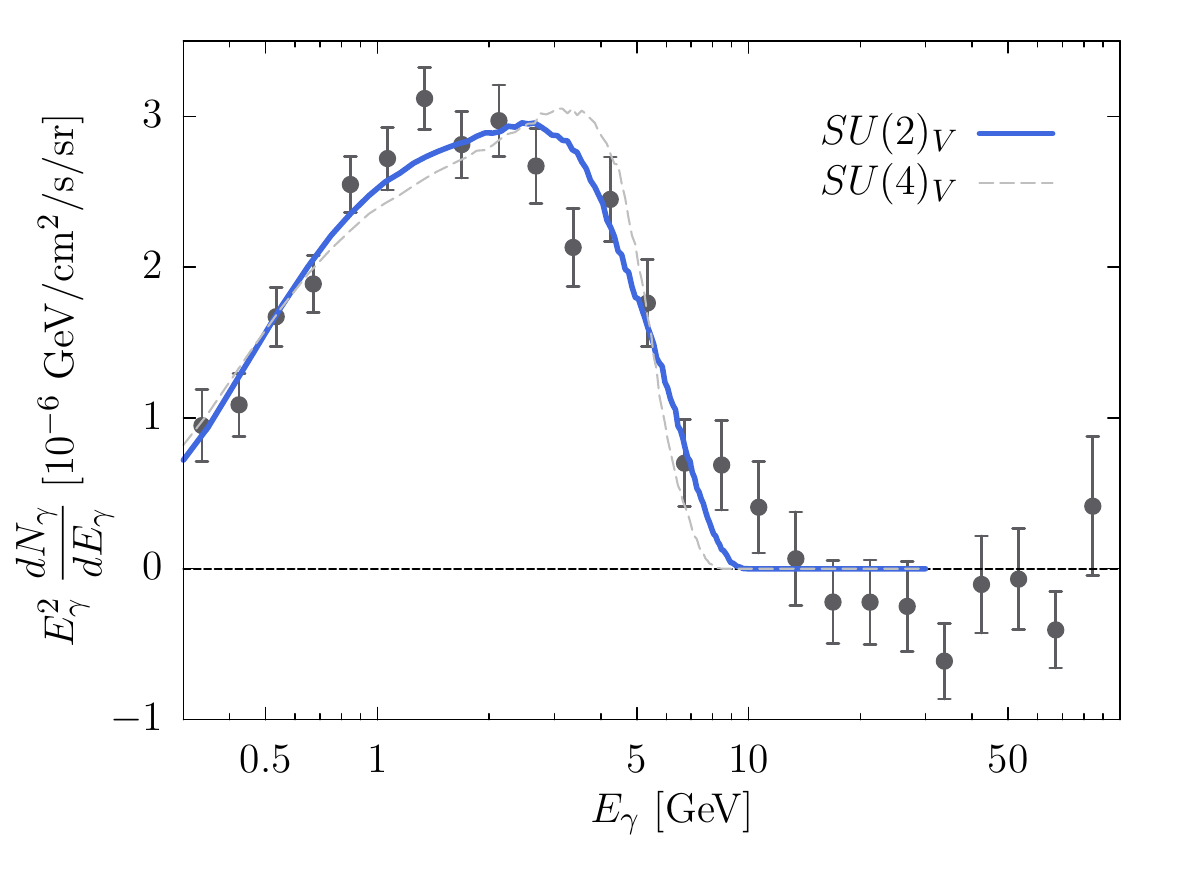}
	\caption{The $\gamma$-ray spectrum at the $SU(2)_{\rm V}$ best-fit point, compared to the $\gamma$-ray excess reported
	in Ref.~\cite{Daylan:2014rsa}. All rates are normalized to the flux at $5^\circ$ from
	the GC, using a generalized NFW halo profile with inner slope $\gamma = 1.26$.
	 }
	\label{fig:SPEC}
\end{figure}

The best-fit $m_\chi$ and $\Lambdav$ for the dark showering spectrum are controlled dominantly by the spectral peak, and are therefore not expected to be modified significantly compared to the results of a full signal plus background analysis. The $SU(2)_{\rm{V}}$ best-fit point corresponds to $m_\chi \simeq 10$~GeV and $\Lambdav \simeq 1$~GeV, with $\chi^2/\rm{dof} = 31.0/24$. This best-fit simulated spectrum is shown in Fig.~\ref{fig:SPEC}, compared to the data of Ref.~\cite{Daylan:2014rsa}. As a point of reference, the best fit for DM annihilating to $b\bar{b}$, with $m_\text{DM} = 32.25$~GeV, gives $\chi^2/\rm{dof} = 29.1/24$ using our methods. 

\begin{figure*}[t]
	\includegraphics[width = 7cm]{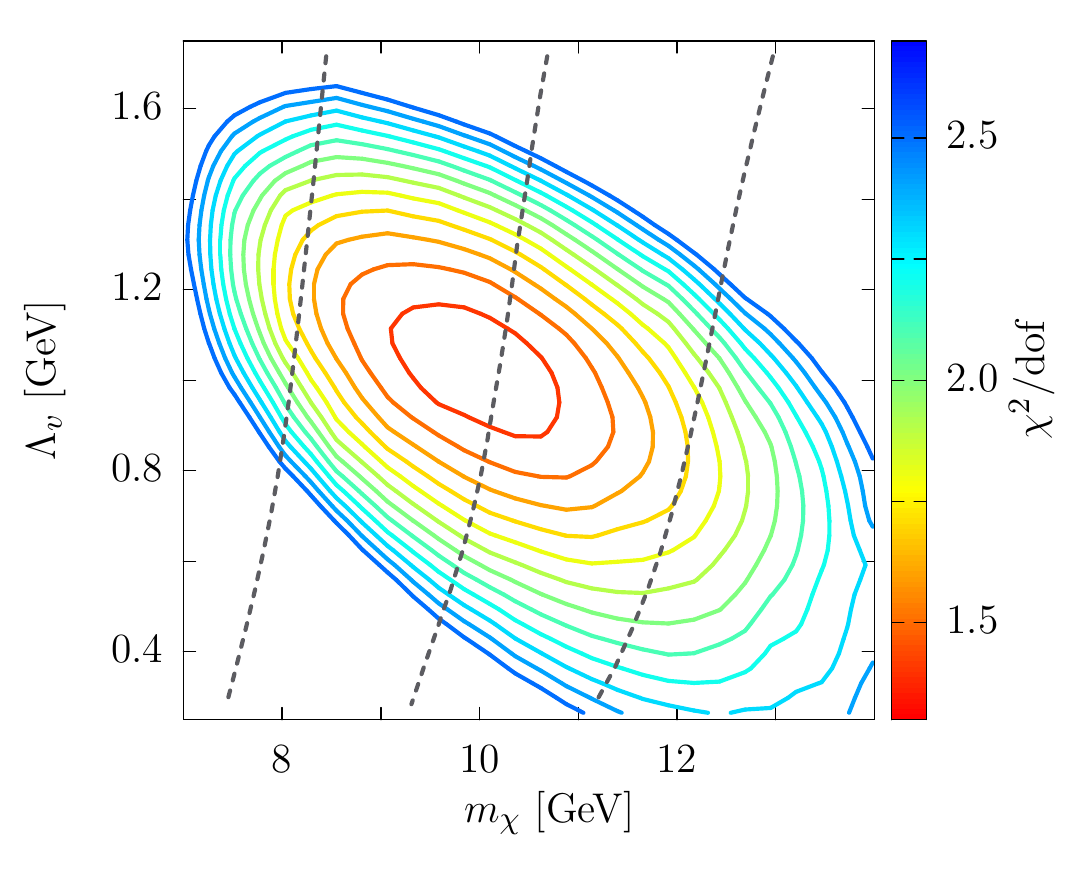}
	\qquad
	\includegraphics[width = 7cm]{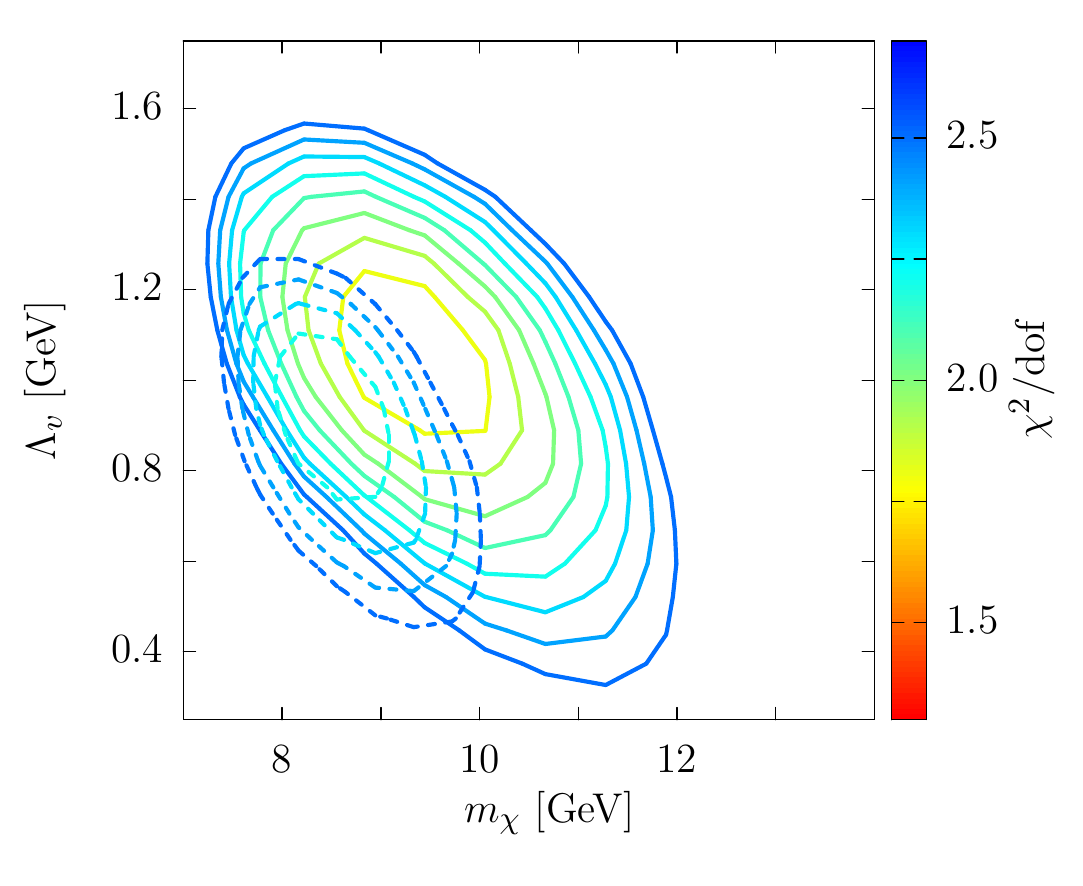}
	\caption{Goodness-of-fit contour heat maps for production of the GC $\gamma$-ray
	excess by a hidden valley shower with gauge groups $\SUV$, $N = 2$ (left)  and $N=3,4$ (right
	solid and dashed respectively).  The cross-section is allowed to float at each point.
	$N=2$ annihilation cross-section contours are also shown for $\langle \sigma v \rangle = 6,8,10
	\times 10^{-27}$~cm$^{-3}$s$^{-1}$ (gray dashed; left to right respectively). The best fit
	occurs at $N=2$, $m_\chi \simeq 10$~GeV, and $\Lambdav \simeq 1$~GeV, with
	a cross-section $\langle \sigma v \rangle =  7.8 \times 10^{-27}$~cm$^{-3}$s$^{-1}$.}
	\label{fig:GOF}
\end{figure*}

Figure~\ref{fig:GOF} displays goodness-of-fit contours over the $m_{\chi}$-$\Lambdav$ plane for $\SUV$, $N = 2,3,4$. One sees that $SU(2)_{\rm{V}}$ is preferred, with relative goodness-of-fit markedly decreasing as $N$ increases. This behavior can be understood as a consequence of the increasing magnitude of the pure gauge contribution to the $\SUV$ $\beta$ function at one loop, which is proportional to $N$. As the $\SUV$ coupling runs faster, the energy window between weak coupling and hadronization -- the energy range over which the parton shower produces high-multiplicity states -- is reduced. This leads to a shorter high-energy tail in the photon energy spectrum and a poorer match to the observed GC spectrum, to the extent that the long tail is a robust feature of the signal data under variation of the DM template.

\textbf{Hidden valley framework.} We now present a hidden valley framework that may realize the above spectra. We focus on a hidden sector charged under $\SUV \otimes \UV$ gauge groups, and millicharged under $\UY$ hypercharge.\footnote{The $\UV$ and $\UY$ gauge fields may also kinetically mix. Under this mixing, milli-electromagnetic charges cannot be generated for SM sterile hidden quarks. Moreover, only transverse modes of the gauge bosons are mixed, so that hidden (pseudo)scalar decay amplitudes via longitudinal $Z$ and $Z'$ modes cancel (see e.g. \cite{Essig:2009nc}). Kinetic mixing therefore mediates hidden (pseudo)scalar decays only at higher loop order, through the $\UV$ chiral anomaly, producing lifetimes that far exceed the Hubble time. We assume hereafter a negligible kinetic mixing compared to the millicharge.} To the extent that millicharged models may be inconsistent with black hole physics \cite{Banks:2010zn}, we consider this model to be merely a phenomenological shorthand for a consistent UV completion.

The matter content of the theory is shown in Table \ref{tab:HVS} below. Here $\uv$ and $\bv$ are hidden quarks, $\chi$ is the DM, $\phi$ is a hidden scalar, and $\lambda_{i}$ can be thought of hidden leptons. The $\lambda_i$ are needed to cancel the $\UV$ and $\UY$ mixed anomalies. Following previous literature \cite{Strassler:2006im}, we call the hidden quarks and hidden leptons ``\emph{$v$-quarks}'' and ``\emph{$v$-leptons}'' respectively, and collectively ``\emph{$v$-fermions}''.  The $v$-quarks and DM are chiral by construction, i.e. $\alpha \not= \pm \beta$ and $\alpha' \not=-\beta'$. The $\lambda_i$ are all $\SUV$ singlets, but may be (milli)charged under $\UV$ ($\UY$). For the sake of brevity, we omit specific charge assignments for $\lambda_i$ in Table \ref{tab:HVS}. Instead, in an appendix we provide an example of a full chiral theory. 

We treat this theory as an effective low energy theory below a scale $M$, and suppose that the scalar $\phi$ has a non-trivial vacuum. The associated vacuum expectation value (VEV) is naturally $\langle \phi \rangle \sim M$. Hence the heavy gauge boson, $Z'_\mu$, associated with $\UV$ breaking,  has mass $m_{Z'} \sim g_v M$, where $g_v$ is the $\UV$ coupling.

\def\arraystretch{1.5}
\begin{table}[tb]
\newcolumntype{C}{ >{\centering\arraybackslash $} m{0.75cm} <{$}}
\newcolumntype{D}{ >{\arraybackslash $} c <{$}}
\begin{center}
\resizebox{\linewidth}{!}{
\begin{tabular*}{1.1\linewidth}{@{\extracolsep{\fill}}|D|CCCC|CCC|C|}
	\hline
	 \quad & \uL & \bL & \uRc & \bRc &  {\chi}_{L}  & {\chi}_R^c & \phi &\lambda_i \\
	\hline\hline
	\SUV & \fund & \fund & \afund & \afund & \id & \id & \id & \id \\
	\UV &  \alpha & -\alpha & \beta & -\beta &  \alpha' & \beta' & \alpha + \beta& \ldots \\
	\UY & \varepsilon & \varepsilon & -\varepsilon & -\varepsilon & 0 & 0 & 0  & \ldots\\
	\hline
\end{tabular*}
}
\end{center}
\caption{Hidden valley charges for left-handed $v$-quarks, DM candidates, and the hidden scalar. Here the hypercharge $\varepsilon \ll 1$ and we require $\alpha' + \beta' = \alpha + \beta$. See appendix for an example of a class of theories with chiral $\lambda_i$ charge assignments.}
\label{tab:HVS}
\end{table}

\emph{Flavor structure --}The hidden valley theory has Lagrangian
\begin{multline}
	\mathcal{L}_{\rm hv}  \supset  y_{\uv} \phi^\dagger \uL \uRc + y_{\bv} \phi \bL \bRc  +  y_\chi \phi^\dagger \chi_R^c \chi_L\\
		+ \lambda_i \lambda_j~\mbox{Yukawas}~. \label{eqn:LHV}
\end{multline}
The $\{\uv, \bv, \chi\}$ basis is explicitly the $v$-quark and DM mass basis of this theory. The Dirac masses are correspondingly $m = y \langle \phi \rangle \sim y M$. The $\lambda_i$, in contrast, may have a richer mass term structure, possibly producing both Dirac and Majorana states. We assume the following mass spectrum
\begin{equation}
	\label{eqn:MSHV}
	m_{\uv} \ll  m_{\chi} \ll M~, \quad   m_{\lambda}, m_{\bv} \lesssim M.
\end{equation}
This spectrum implies that non-relativistic $\chi$ may only annihilate into $\uv\bar{\uv}$. As will be discussed further below, only $\chi$ will produce a significant thermal relic.

The $\UV$ charge assignments ensure that the maximal flavor symmetry in the $v$-quark kinetic terms is $U(1)^4$. This chiral symmetry is broken by the yukawas to $U(1)^{2}$: independent $U(1)$ hidden baryon symmetries therefore survive for the $\uv$ and $\bv$ $v$-quarks respectively.

\emph{Light pseudoscalar --} Now let $\SUV$ be confining at a scale $\Lambdav \ll M$, such that $\uv$ quarks remain light. That is,
\begin{equation}
	\label{eqn:SLQF}
	 m_{\uv} \ll \Lambdav \ll m_{\bv}~.
 \end{equation}
For the present case of a single light $v$-quark flavor \eqref{eqn:SLQF}, this confinement breaks the axial $U(1)$ $\uv$-flavor symmetry. However, the $\SUV$ instanton with this axial $U(1)$ ensures that there is no light pseudogoldstone boson in the spectrum \cite{Witten:1979vv}. Instead, the light pseudoscalar is $\etav \sim \uv\overline{\uv}$, analogous to the QCD pseudoscalar $\eta'$: we shall refer to it as the hidden eta. Similarly, we shall label $v$-quark hadrons by their QCD equivalents, adding a `$v$' subscript. In the large $N$ limit, the $\etav$ has mass
\begin{equation}
	\label{eqn:PM}
	m_{\etav} \sim \Lambdav/\sqrt{N}~.
\end{equation}

\emph{Benchmark parameters --} To focus the discussion, hereafter we parametrize the analysis around a neighborhood of the benchmark values 
\begin{gather}
	\varepsilon^2  =10^{-7}~,  \quad  m_{\etav}   = 500~\mbox{MeV}~, \quad M  =450~\mbox{GeV}~, \notag\\  m_\chi = 10~\mbox{GeV}~, \quad \mbox{and}  \quad N = 2~. \label{eqn:BPV}
\end{gather}
For these parameters, the $\etav$ mass relation \eqref{eqn:PM} implies a benchmark $\Lambdav \lesssim 1$~GeV. The choices for $N$, $\Lambdav$ and $m_\chi$ are motivated by the best fit regions of the showering simulation shown above in Fig.~\ref{fig:GOF}. These choices also anticipate various astrophysical, cosmological and other empirical bounds.

\emph{Bound state spectrum --} The one-light-flavor bound state spectrum contains the vector bound state $\omega_v \sim \uv\overline{\uv}$. The hidden $\uv$ baryon number ensures that the lightest hidden baryon, $\Delta_v \sim N \uv$, is stable. The masses of the $\omega_v$ and $\Delta_v$ are typically
\begin{equation}
	m_{\omega_v} \sim \Lambdav~,\qquad m_{\Delta_v} \sim N \Lambdav~.
\end{equation}
For $N=2$, $\Delta_v$ and all other bound states are bosons. However, for the sake of generality, we will also consider the possibility of fermionic bound states in the discussion that follows. The analysis below may be suitably modified for fermionic $\Delta_v$s, with similar conclusions.

Along with the $\etav$ and $\omega_v$ one expects parity conjugate bound states, namely the scalar meson $\sigmav$ and pseudovector ${h_1}_v$, as well as higher spin states. Typically $m_{\sigmav, {h_1}_v} \gtrsim m_{\etav, \omega_v}$ respectively. Orientifold planar equivalence arguments \cite{Armoni:2005qr,Sannino:2003xe} imply that for large $N$ one expects $m_{\etav}/m_{\sigmav} \simeq 1 - 2/N$. Numerical lattice simulations \cite{Demmouche:2008ms} (\cite{Armoni:2008nq,Farchioni:2008na}) support $m_{\sigmav} > 2m_{\etav}$ for the case of $N=2$ ($N=3$). Hence the $\sigmav$, and presumably all other higher spin states, may be broad for $N$ not too large. We will, however, also contemplate below the phenomenology of a narrow $\sigmav$, in order to cover all possibilities.

\begin{figure}[t]
	\includegraphics[width = 0.95\linewidth]{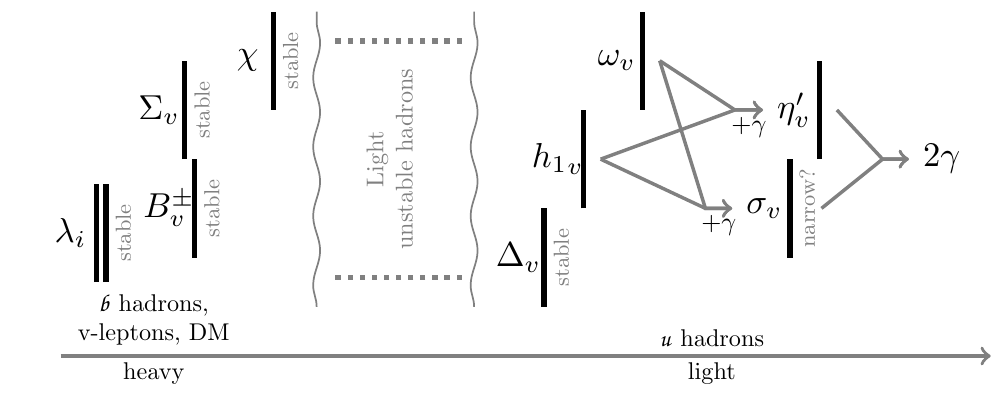}
	\caption{Speculative spectrum of DM, $v$-leptons, light (meta)stable and heavy bound states. Certain dominant decay modes are shown with grey arrows.}
	\label{fig:SBS}
\end{figure}

Because of the independent $\uv$ and $\bv$ baryon symmetries, there are also heavier stable $\bv$ hadrons in the spectrum, for example $B^+_v \sim \uv \bar{\bv}$ or $\Sigma_v \sim \uv^{N-1}\bv$, among many others. The heavy $\bv$ mass \eqref{eqn:MSHV} ensures that these hadrons are not produced in DM annihilations. The spectrum of bound states, $v$-leptons and DM is summarized in Fig.~\ref{fig:SBS}. 

\emph{$\etav$ decay modes -- } The $\etav$ decays to photons through the hypercharge chiral anomaly operator $\etav F_{\mu\nu}\tilde{F}^{\mu\nu}$. Including the effects of millicharges, the decay rate is
\begin{equation}
	\label{eqn:PGGR}
	\Gamma[\etav \to 2 \gamma]  \sim \frac{N^2\alpha^2 \varepsilon^4}{36 \pi} \frac{m_{\etav}^3}{\Lambdav^2}~ ,
\end{equation}
where $\alpha \equiv e^2/4\pi$ is the fine structure constant.  In the spirit of naive dimensional analysis (NDA), we have also employed the approximate relation $f_{\etav} \sim \Lambdav/4\pi$, where $f_{\etav}$ is the $\etav$ decay constant.

The dominant leptonic decay mode of the $\etav$ is the Dalitz process $\etav \to \gamma^* (\to \ell^+\ell^-)\gamma$. Just as for the SM $\pi^0$, the $\etav \to \ell^+\ell^-\gamma$ rate is suppressed approximately by a factor of $e^2$ and phase space factors with respect to $\etav \to 2\gamma$. Hidden eta decay modes to other SM species dominantly occur through the $\etav F \tilde{F}$ operator, too: Albeit $1/m_Z^2$-suppressed decays through a longitudinal $Z$ do not occur because the $\uv$ $v$-quark has zero axial coupling to hypercharge in this HV theory.  That is, direct decays to $\ell^+\ell^-$ occur via loop amplitudes of form
\begin{equation}
	\label{eqn:ELL}
	\mathcal{M}_{\etav \to \ell^+\ell^-} = \parbox{4cm}{\includegraphics[scale =0.8]{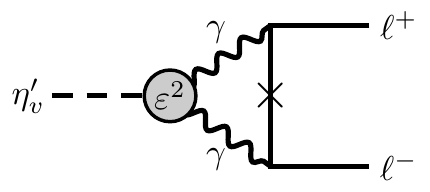}}~,
\end{equation}
and are therefore helicity and electroweak loop suppressed. Away from the threshold regime, $1 - 2m_\ell/m_{\etav} \ll 1$, one estimates \cite{Dorokhov:2007bd},
\begin{equation}
	\frac{\Gamma[\etav \to \ell^+\ell^-]}{\Gamma[\etav \to 2\gamma]} \sim 2\alpha^2\frac{m^2_\ell}{m^2_{\etav}}\log^2\bigg(\frac{m_\ell}{m_{\etav}}\bigg)~,
\end{equation}
which has an upper bound of $\alpha^2$ for any charged fermion SM species. Parity conserving $\etav$ decays to charged scalar SM species feature only electroweak loop suppression (cf. \eqref{eqn:ELL}) compared to the diphoton rate. All these rates therefore remain small, even compared to the Dalitz $\etav \to \ell^+\ell^-\gamma$ rates.

\emph{Narrow $\sigmav$ --} If $m_{\sigmav} < 2m_{\etav}$, the mode $\sigmav \to 2 \etav$  is kinematically forbidden, rendering $\sigmav$ with a narrow width. However, the scalar $\sigmav$ may decay to photons via a $\Delta_v$ loop, with rate $\Gamma_{\sigmav}/\Gamma_{\etav} \sim (m_{\sigmav}/m_{\etav})^3$. Couplings of $\sigmav$ to SM species through a longitudinal $Z$ are not permitted by parity, and there is no scalar messenger between the HV and SM sectors. Hence, $\sigmav$ decays to fermionic (scalar) SM species must also proceed through photon loops or Dalitz processes, as in eq. \eqref{eqn:ELL}, and are helicity and electroweak loop (electroweak loop) suppressed compared to the photon rate.  

In summary, the $\etav$ and narrow $\sigmav$ lifetimes, together with their coupling to SM species, are similar. This implies that narrow $\sigmav$'s may be treated as $\etav$'s not only within the hidden valley shower to photons, as they have been above, but also within cosmological, astrophysical and collider contexts, to be discussed below.

\emph{Other decay modes --} The absence of C violation in this theory forbids the $\omega_v$ to decay into any combination of $\etav$ and $\sigmav$ alone. Instead only decays such as $\omega_v \to \etav + \mbox{SM}$ are permitted. By comparison to the SM, we expect $\omega_v \to \etav \gamma$ to be the dominant $\omega_v$ decay mode, with subdominant $\omega_v \to \gamma^* \to \ell^+\ell^-$.  Up to hadronic uncertainty and p-wave suppression, we expect $\omega_v$ to have a width $\Gamma_{\omega_v} \sim \varepsilon^2 \alpha m_{\omega_v}$. The corresponding lifetime $\tau_{\omega_v} \lesssim 10^{-15}$~s. The mode ${h_1}_v \to \etav \gamma$ is expected to proceed comparatively faster, due to absence of p-wave suppression and a larger phase space. The dominant metastable state decay modes are inventoried in Fig.~\ref{fig:SBS}. 

\textbf{Thermal history.} \emph{Hidden baryon abundances --} In the confined phase of the hidden valley, i.e. $T \lesssim \Lambdav$, the hidden baryons $\Delta_v$ are coupled to the $\etav$'s via marginal and higher-dimension operators that encode the strong dynamics. Freeze-out of the $\Delta_v$'s is controlled by operators such as
\begin{equation}
	\label{eqn:CPTHO}
	\etav^2 \Delta_v \Delta_v^c~, \quad \mbox{or} \quad \frac{1}{\Lambdav^2} {\omega_v}_{\mu\nu}{\omega_v}^{\mu\nu} \Delta_v \Delta_v^c~,
\end{equation}
with thermally averaged cross-section
\begin{equation}
	\langle \sigma v \rangle_{\Delta_v} \sim \bigg[\frac{2}{N}\bigg] \bigg[\frac{500~ \rm{MeV}}{m_{\etav}}\bigg]^26\times10^{-17} ~\mbox{cm}^3\mbox{s}^{-1}~,
\end{equation}
assuming a freeze-out temperature $x_f \equiv m_{\Delta_v}/T_f \sim 20$. The result is similar for a fermionic $\Delta_v$. This is very large compared to the typical WIMP-type cross-section $\langle \sigma v\rangle \sim 10^{-26}~\mbox{cm}^3\mbox{s}^{-1}$. Assuming no hidden baryon asymmetry in the HV sector, the thermal relic abundance of hidden baryons is therefore negligible. The same analysis applies \emph{a fortiori} to the much heavier $B^\pm_v$, $\Sigma_v$ and other stable $\bv$ hadrons.

\emph{Dark matter production --} The Dirac field $\chi$ couples at tree-level only to the HV degrees of freedom via the $Z'$. Our assumed spectrum \eqref{eqn:MSHV} implies that its dominant annihilation channel is into the light $\uv$ quarks through the $Z'$. The cross-section is dominated by the s-wave channel, with the explicit form
\begin{equation}
	\label{eqn:SVCC}
	\langle\sigma v\rangle_{\bar{\chi}\chi\to\bar{\uv}\uv} =\frac{m_{\chi}^2\sqrt{1-\hat{m}^2}}{\pi\langle\phi\rangle^4}\bigg[1+\frac{\hat{m}^2}{2}\bigg] \sim \frac{m_\chi^2}{\pi M^4}~,
\end{equation}
where $\hat{m}\equiv m_{\uv}/m_{\chi} \ll 1$, and $m_{Z'}^2=g_{v}^2\langle\phi\rangle^2 \sim g^2_v M^2$ is used. 

From the prompt photon spectra, presented in Figs~\ref{fig:GOF}, the $N=2$ best fit annihilation cross-section is
\begin{equation}
	\langle \sigma v \rangle_{\bar{\chi}\chi\to\bar{\uv}\uv} \simeq 7.8\times 10^{-27} \mbox{cm}^3\mbox{s}^{-1}~.
\end{equation}
For the benchmark $m_\chi = 10$~GeV, one then obtains $M \sim 450$~GeV, motivating our benchmark choice for this parameter.  Since there are no other significant $\chi$ annihilation channels, from this s-wave annihilation cross-section one may estimate the surviving relic density, viz.
\begin{equation}
	[\Omega h^2]_{\chi} \sim 0.3 \bigg[\frac{x_f}{20}\bigg] \bigg[\frac{8\times 10^{-27}~\text{cm}^3\text{s}^{-1}}{ \langle \sigma v \rangle_{\bar{\chi}\chi\to\bar{\uv}\uv}} \bigg] ~.
\end{equation}
Up to the $\mathcal{O}(1)$ uncertainties inherent in this discussion, this is in agreement with the observed relic density of dark matter.

\emph{Hidden lepton abundances --}  Freeze out of the $v$-leptons, $\lambda_i$, is similarly controlled by $\lambda \lambda \to \bar{\uv}\uv$ via the $Z'$. From the assumed mass spectrum \eqref{eqn:MSHV}, if the $v$-leptons have typical mass $m_{\lambda_i} \lesssim m_{Z'}/2$, then the regime of eq. \eqref{eqn:SVCC} applies, with corresponding annihilation cross-section
\begin{equation}
	\label{eqn:CPA}
	\langle \sigma v \rangle_{\lambda_i} \sim \bigg[ \frac{m_{\lambda_i}}{500~\mbox{GeV}}\bigg]^2 10^{-23}  \mbox{cm}^3\mbox{s}^{-1}~.
\end{equation}
For larger $m_{\lambda_i}$, the $Z’$ may be on-shell, increasing the size of cross-section even further. 
The relic abundance of $v$-leptons is therefore negligible.

\emph{Big-bang nucleosynthesis (BBN) bounds -- } In the free phase of the HV theory, the $v$-fermions are recoupled to the SM plasma via photon exchange below a temperature $T_{\rm rec} \sim \varepsilon^2 \alpha^2 M_{\rm pl}/\sqrt{g_*} \sim 10^5$~TeV, at the benchmark for $\varepsilon$. As a consequence, the hidden quark-gluon plasma exceeds the bounds on the number of relativistic degrees of freedom, $N_{\rm eff}$, during BBN -- the BBN temperature, $T_{\rm BBN} \lesssim 3$~MeV -- unless they are confined. That is, we require $\Lambdav \gg 3$~MeV.

From the point of view of the confined phase of the HV, provided $m_{\etav} \gg T_{\rm BBN}$, the hidden $\etav$s are too heavy for efficient $\etav$ rethermalization by inverse decays at the BBN epoch. There is therefore no direct tension with the BBN $N_{\rm eff}$ or $^4$He bounds. However, if the $\etav$s have a sufficiently long lifetime and decouple with non-negligible abundance, there is a risk of a post-BBN matter-dominated epoch generated by non-equilibrium long-lived $\etav$s, and subsequent significant entropy production once they decay. These potential problems can be avoided, even without investigating their underlying details, by simply requiring that the $\etav$s decay well-before the BBN and neutrino decoupling epoch. 

\begin{figure}[t]
	\includegraphics[width = 0.9\linewidth]{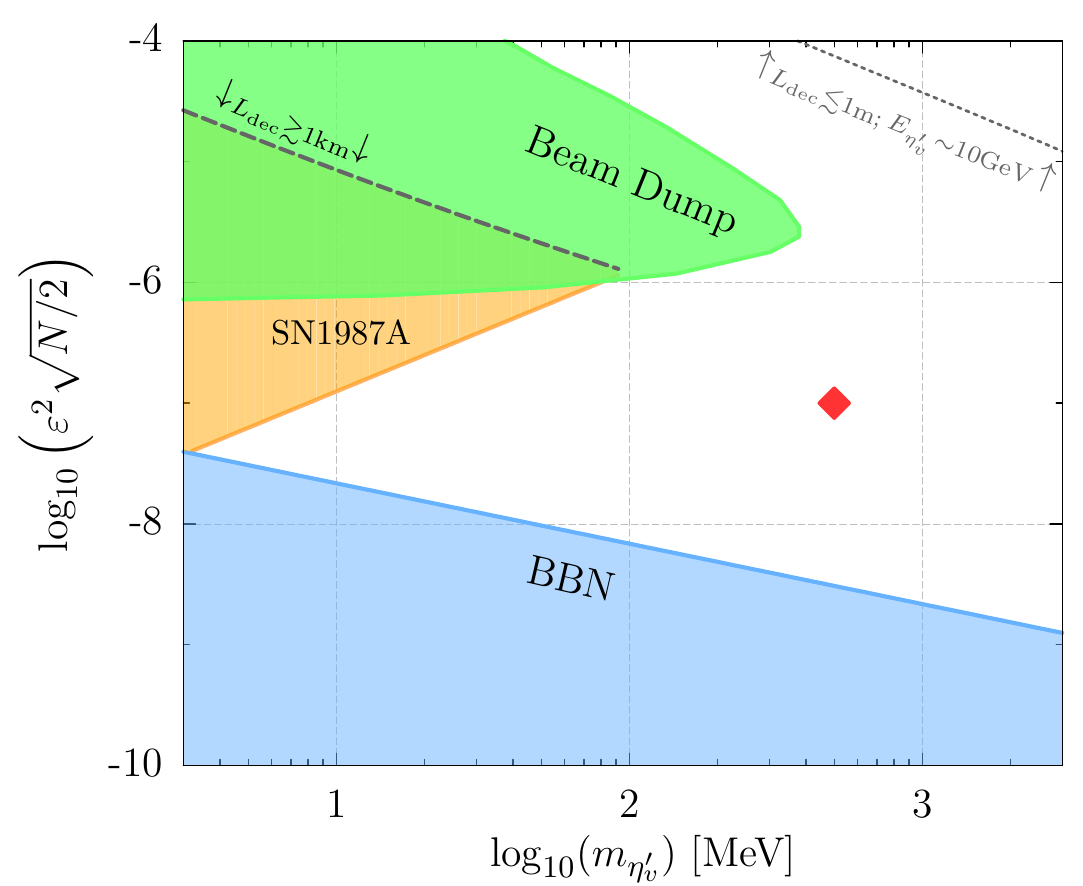}
	\caption{ALP beam dump \cite{Prinz:1998ua} (green), SN1987A \cite{Turner:1987by} (orange; dashed line) and BBN (blue) bounds for the $\etav$, with SN temperature $T \simeq 30$~MeV. Also shown is the displaced vertex region $L_{\rm{dec}} \lesssim 1$~m for photon energies $\sim 10$~GeV (dotted line); and benchmark values \eqref{eqn:BPV} for the $\{m_{\etav} ,\varepsilon, N\}$ parameter space (red diamond).}
	\label{fig:ALPB}
\end{figure}

Taking all these considerations together, to avoid BBN bounds it is sufficient to require
\begin{align}
	\tau_{\etav} & \ll H^{-1}_{\rm{BBN}} \sim 10^{-1}~\mbox{s}~, \label{eqn:LB}\\
	(\Lambdav \gtrsim m_{\etav}) & \gg T_{\rm{BBN}} \sim 3~\mbox{MeV}~.\label{eqn:MB}
\end{align}
Combining eqs. \eqref{eqn:PM}, \eqref{eqn:PGGR} and \eqref{eqn:LB} results in a BBN bound on the $\{m_{\etav},\varepsilon, N\}$ parameter space, shown in Fig.~\ref{fig:ALPB} with several other bounds, to be discussed below.

It is possible that a narrow $\sigmav$ may have a mildly longer lifetime than the $\etav$'s, so that they may still be marginally metastable at the BBN epoch. However, since $\sigmav$'s couple to $\etav$'s via the marginal operator $\etav^2\sigmav^2$, and since generically the splitting $|m_{\sigmav} - m_{\etav}| \sim m_{\etav}$ for $N$ not too large, they annihilate efficiently into $\etav$'s at the temperature epoch $T_{\rm BBN} \ll T \lesssim m_{\etav}$, leaving behind a negligible metastable thermal relic by the BBN epoch (cf. $\Delta_v$ freeze-out).  A similar analysis applies to other, heavier $v$-hadronic BBN metastable states, if any.

\textbf{Axion-like particle (ALP) bounds.} The HV framework admits only suppressed couplings to the SM sector. Nevertheless, several collider and astrophysical searches may still set constraints on the $\{m_{\etav},\varepsilon, N\}$ parameter space. Depending on their energy scale, such searches can either produce $v$-fermions directly or generate $v$-hadrons.

The $\etav$ interacts with the SM sector dominantly via the chiral anomaly operator (cf. eq. \eqref{eqn:PGGR})
\begin{equation}
	\label{eqn:EPC}
	\mathcal{O}_{\gamma\gamma} = \frac{1}{4}g_{\gamma\gamma}\etav F_{\mu\nu}\tilde{F}^{\mu\nu}~,\quad g_{\gamma\gamma} \sim \frac{4\varepsilon^2N\alpha}{3\Lambdav}~.
\end{equation}
The $\etav$ therefore acts as an ALP below the confinement scale, $\Lambdav$. In this operator language, ALP bounds on $g_{\gamma\gamma}$ can set bounds on the HV framework independently of the particular realization of $\etav F \tilde{F}$ coupling.

A general study of the constraints on ALP-type couplings can be found in several works (see e.g. Refs \cite{Davidson:2000hf,Hewett:2012ns}). Relevant constraints around the benchmark parameters are set by the SLAC beam dump experiment \cite{Prinz:1998ua}, SN1987A \cite{Turner:1987by}, and BBN (see e.g. \cite{Cadamuro:2011fd}). However, in contrast to the standard ALP picture, here the $\etav$  coupling to SM fermions may only be generated via photon loop-suppressed amplitudes involving $\mathcal{O}_{\gamma\gamma}$. From na\"\i ve dimensional analysis, one estimates the corresponding operator
\begin{equation}
	\label{eqn:ESMFC}
	\mathcal{O}_{ff} = g_{ff} \etav \bar{f} \gamma^5 f~, \quad g_{ff} \sim g_{\gamma\gamma} m_f \frac{\alpha}{4\pi}~.
\end{equation}

The operator $\mathcal{O}_{ff}$ plays a vital role in the cooling of supernova (SN) cores by the nucleon bremsstrahlung $2n \to 2n\etav$. Typically, there is a forbidden window, conservatively \cite{Raffelt:2006cw}
\begin{equation}
	10^{-10} \lesssim g_{nn} \lesssim 10^{-6}~,
\end{equation}
which may be translated into bounds on $g_{\gamma\gamma}$, and hence on  the $\{m_{\etav}, \varepsilon, N \}$ parameter space, by the relation \eqref{eqn:ESMFC}. There may also be a $2n \to 2n\etav\gamma$ bremsstrahlung contribution to the cooling, also mediated by $g_{\gamma\gamma}$. However, the resulting cross-section is of the similar order to $2n \to 2n\etav$, and so does not change the order of magnitude estimation for the $g_{\gamma\gamma}$ bounds. 

If the $\etav$ mass is above $100$ MeV, its production rate inside the SN core is Boltzmann suppressed, and is then too low to set a significant constraint on $\varepsilon$. Alternatively, the SN bounds are lifted within the regime that that the $\etav$ decay length, $L_{\rm dec} = \gamma\beta c \tau_{\etav}$, is much smaller than the supernova core radius $\sim10$~km, i.e. if $L_{\rm dec} \lesssim 1$~km. Noting that the $\etav$ energy $ \sim  m_{\etav} + 3 T/2$, from eq. \eqref{eqn:PGGR} this regime is explicitly
\begin{equation}
	\label{eqn:DLR}
	\varepsilon^2\bigg[\frac{N}{2}\bigg]^{1/2} \!\!\!\! \gtrsim 10^{-6}\bigg(\!\frac{3 T}{m_{\etav}} + \frac{9}{4}\frac{T^2}{m_{\etav}^2}\!\bigg)^{1/4}\bigg[\frac{100~\rm{MeV}}{m_{\etav}}\bigg]^{1/2}.
\end{equation}
The ALP \eqref{eqn:EPC}--\eqref{eqn:DLR} and BBN \eqref{eqn:LB}--\eqref{eqn:MB} bounds are summarized in Fig.~\ref{fig:ALPB}. One sees that a neighborhood of the benchmark values \eqref{eqn:BPV} is unconstrained, but may be probed by future beam dump experiments.

\textbf{Terrestrial constraints.} \emph{Collider searches -- } A generic millicharged particle has a long scattering length compared to the SM leptons in a detector. For example, the lead tungstate (PbWO$_4$) crystals of the electromagnetic calorimeter at CMS has a thickness of $\simeq 25$ radiation lengths \cite{Bayatian:922757}. A $v$-fermion with $\varepsilon<0.1$ has a scattering length that is $\sim10^2$ times longer than an electron, and so does not scatter inside the crystal. We may therefore treat $v$-fermions as missing energy in the context of collider searches. 

In a typical missing energy search such as jet$+$MET \cite{Khachatryan:2014rra,ATLAS-CONF-2012-147}, the cross section at LHC $8(14)$ TeV  is of the size
\begin{equation}
	\sigma_{pp\to j\uv\bar{\uv}}\simeq 2(7)\times 10^{-5}\bigg[\frac{\varepsilon^2}{10^{-7}}\bigg]{\rm fb}~,
\end{equation}
for the leading jet cuts $p_T>110$ GeV, $|\eta_j|<2.4$, and $\slashed E_T>250$ GeV. The actual $8$ TeV CMS mono-jet \cite{Khachatryan:2014rra} only excludes $\varepsilon>0.6$ at $90\%$ confidence when assuming invisible $v$-fermions. Similarly, the mono-$\gamma$ search at LEPII \cite{Abdallah:2008aa} has cross section
\begin{equation}
\sigma_{e^+e^-\to \gamma\uv\bar{\uv}}\simeq 2\times 10^{-4}\bigg[\frac{\varepsilon^2}{10^{-7}}\bigg]{\rm fb},
\end{equation}
for center of mass energy $200$ GeV, photon energy cut $E_{\gamma}>6$ GeV and polar angle cut $45^0<\theta<135^0$.  The region $\varepsilon^2 >10^{-5}$ may be probed at a high luminosity LHC run, but otherwise both of these experiments are insensitive to the benchmark millicharge. 

Besides the production of $v$-fermions, one can consider searches for the $Z'$ of U$(1)_V$. However, the $Z'$ in this model only couples to visible sector through loop-level vector boson mixing, and the corresponding $Z'$ search sets a looser constraint on the model than direct $v$-fermion production. Similarly, the SM $Z$ boson can decay into $\uv\bar{\uv}$, but the partial width is of order $\simeq 10^{-5}(\varepsilon^2/10^{-7})$ MeV, which is negligible compared to the precision of the current $Z$ width measurement.

\emph{Displaced vertices -- } Although $v$-fermions act like missing energy within collider searches, the $\etav$s showered by $v$-fermion hadronization may decay to photons inside the detector. The dotted line in Fig.~\ref{fig:ALPB} shows a lower bound for $L_{\rm dec} \lesssim 1$~m and photon energies around $10$~GeV. Future searches of the soft and displaced photons plus missing energy can provide interesting constraints on the HV model, if its parameters are near this lower bound.

\emph{Direct detection --} Scattering between the DM $\chi$ and nuclei can be mediated by a loop-induced $Z'$-photon mixing, such as
\begin{equation}
	\parbox{5.25cm}{\includegraphics[scale =0.8]{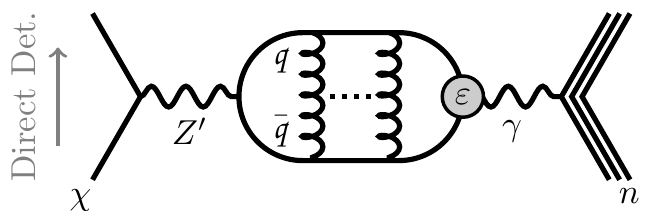}}~.
\end{equation}
Compared to WIMP-type scattering, the additional $\varepsilon^2\alpha^2$ and $m_{Z'}$ suppression produces a cross-section smaller than $\sim 10^{-49} (\varepsilon^2/10^{-7})$ cm$^2$ for the benchmark values.  This is hidden below the direct detection neutrino floor.

\textbf{Alternative HV scenarios.} Finally, having discussed the phenomenology of the millicharge theory in detail, various alternative scenarios may be entertained~\cite{Fan:2012gr}. For example, one can conceive of $\etav$ mixing with an electromagnetic axion, that mediates decays to photons. Another attractive possibility is mediation of the $\etav$ diphoton coupling through a heavy $v$-quark, $\mathpzc{t}$, with unit hypercharge. Typically the effective millicharge coupling scales as $\varepsilon^2 \sim \Lambdav^4/(16 \pi^2 m_{\mathpzc{t}}^4) \lesssim 10^{-10}$ from collider chargino bounds. This is in tension with the BBN bound \eqref{eqn:LB}. Construction of these types of HV scenarios likely requires more elaborate theories than the simplest cases, as well as a careful treatment of their cosmological implications, should the bounds \eqref{eqn:LB}--\eqref{eqn:MB} no longer be applicable.

\textbf{Conclusions.} We have presented a class of hidden valley models that may produce the observed galactic center $\gamma$-ray excess through the dark showering and subsequent electromagnetic decay of dark hadrons. Compared to the more conventional picture of DM annihilation into SM $b$ quarks or $\tau$ leptons, this scenario permits production of the $\gamma$-ray signal without associated antiproton or positron fluxes. Moreover, it permits the SM and DM sectors to be coupled far more weakly than limits placed by current cosmological, astrophysical, and collider constraints, while still producing the observed DM abundance. 

In the minimal setup, for which the parton shower parameters are determined by the confinement scale $\Lambda_v$ and DM mass $m_{\chi}$, we obtain the best fit of the \emph{Fermi}-LAT data with a hidden $SU(2)_\text{V}$ confining group, $m_{\chi} \simeq 10$~GeV and $\Lambdav \simeq 1$~GeV. The corresponding annihilation cross section is $\langle\sigma v\rangle=8\times 10^{-27}$ cm$^3/$s. In general, we find lower rank gauge groups produce a longer tail of high-energy photons, because of the slower running of the coupling. To the extent that a long tail is a robust feature of the signal data under variation of the DM template, this latter result is an \emph{a posteriori} motivation for considering showering by (dark) confining groups other than the SM $SU(3)$ color group.

Although the hidden valley setup can relax antiproton and positron bounds, there are still non-trivial model building challenges that must be met in order to construct a realization of the framework that has both dominant dark hadron coupling to photons, as well as a short enough lifetime to evade BBN constraints. Here we have presented a generic class of millicharge models that have the required dominant dark hadron-photon coupling, realistic thermal histories, and satisfy pertinent cosmological, astrophysical and collider bounds. The bounds from various ALP searches, including SN cooling, beam dump experiments, and BBN constraints exclude part of the $\{m_{\etav},\varepsilon,N\}$ parameter space, in a model-independent manner, leaving a large region unconstrained. Future improvements of the beam dump search in particular can further constrain this parameter space.

Finally, other realizations of the hidden valley framework may also exist. For example, models with axionic or heavy quark mediation between the SM and hidden valley sectors.

\textbf{Acknowledgements.}
The authors thank Yang Bai, Joshua Berger, Thomas Dumitrescu, Ben Heidenreich, Simon Knapen, Duccio Pappadopulo, Michele Papucci, Matthew Reece, Daniel Stolarski, Philip Tanedo, Jon Walsh, and Kathryn Zurek for helpful discussions. The work of MF is supported by the Department of Energy (DoE) under grant DE-SC003916 and the National Science Foundation (NSF) under grant No.~PHY-1258729. The work of DR is supported by the NSF under grant No.~PHY-1002399. The work of YT is supported by the DoE under Grant DE-FG02-91ER40674. This work was also supported in part by the NSF under grant No. PHYS-1066293 and the hospitality of the Aspen Center for Physics.

\appendix
\textbf{Appendix.} In this appendix, we present an example of a fully chiral millicharged theory, which may realize the partial theory of Table \ref{tab:HVS} above. The $\UV$ charge assignments shown are a specific case of a much more general class of possibilities.
\def\arraystretch{1.5}
\begin{table}[ht]
\newcolumntype{C}{ >{\centering\arraybackslash $} m{0.7cm} <{$}}
\newcolumntype{E}{ >{\centering\arraybackslash $} m{1cm} <{$}}
\newcolumntype{D}{ >{\arraybackslash $} c <{$}}
\begin{center}
\resizebox{\linewidth}{!}{
\begin{tabular*}{1.1\linewidth}{@{\extracolsep{\fill}}|D|CCCC|EEEE|}
	\hline
	 \quad & \uL & \bL & \uRc & \bRc &  \xL & \zL & \xRc & \zRc \\
	\hline\hline
	\SUV & \fund & \fund & \afund & \afund & \id & \id & \id & \id \\
	\UV & 1 & -1 & \frac{3}{5} & -\frac{3}{5} &  1 + \frac{5}{4} & \frac{5}{4} -1 & \frac{3}{5}  - \frac{5}{4} & -\frac{3}{5} -\frac{5}{4} \\
	\UY & \varepsilon & \varepsilon & -\varepsilon & -\varepsilon &  -\varepsilon N & -\varepsilon N & \varepsilon N  & \varepsilon N\\
	\hline
	\hline
	 \quad & {\chi}_{L}  & {\chi}_R^c & {\lambda_3}_L & {\lambda_3}_R^c  & \phi &&& \\
	\hline\hline
	\SUV & \id & \id & \id & \id & \id &&& \\
	\UV & \frac{1}{5}  & -\frac{9}{5} & \frac{4}{5} & \frac{4}{5} &  \frac{8}{5} &&&  \\
	\UY & 0 & 0&0 & 0 & 0 &&& \\
	\hline
\end{tabular*}
}
\end{center}
\caption{Hidden valley charges for left-handed $v$-quarks,  $v$-leptons, DM candidates, and the hidden scalar. Here the hypercharge $\varepsilon \ll 1$.}
\end{table}
This hidden valley theory is manifestly non-anomalous, and has Lagrangian
\begin{multline}
	\mathcal{L}_{\rm hv}  \supset  y_{\uv} \phi^\dagger \uL \uRc + y_{\bv} \phi \bL \bRc  +  y_{1} \phi^\dagger \xL \xRc + y_{2} \phi \zL \zRc\\
		 + y_{3}^{LL} \phi {\lambda_3}_L {\lambda_3}_L + y^{LR}_{3} \phi {\lambda_3}_R^c  {\lambda_3}_L + y^{RR}_{3} {\lambda_3}_R^c  {\lambda_3}_R^c  \\
		 + y_\chi \phi^\dagger \chi_R^c \chi_L ~.
\end{multline}
The $v$-leptons $\lambda_{1,2}$ gain only Dirac masses under $\UV$ breaking by $\langle \phi \rangle$, while $\lambda_3$ has Majorana mass terms, and produces two Majorana states.

\bibliography{HiddenValley}

\end{document}